\documentstyle[12pt,aasms4]{article}

\received{1999-NOV-30}
\accepted{2000-Jan-7}
\slugcomment{to appear in {\it The Astrophysical Journal}}

\lefthead{Kraemer, George, Turner, \& Crenshaw}
\righthead{On The Reddening in X-ray Absorbed Seyfert 1 Galaxies}

\begin{document}

\title{On The Reddening in X-ray Absorbed Seyfert 1 Galaxies}

\author{S. B. Kraemer\altaffilmark{1},
I. M. George\altaffilmark{2},
T. J. Turner\altaffilmark{3},
\& D. M. Crenshaw\altaffilmark{1}}

\altaffiltext{1}{Catholic University of America,
NASA's Goddard Space Flight Center, Code 681,
Greenbelt, MD  20771; stiskraemer@yancey.gsfc.nasa.gov, 
crenshaw@buckeye.gsfc.nasa.gov.}

\altaffiltext{2}{Universities Space Research Association,
NASA's Goddard Space Flight Center, Code 660,
Greenbelt, MD  20771; ian.george@gsfc.nasa.gov.}

\altaffiltext{3}{University of Maryland, Baltimore County,
NASA's Goddard Space Flight Center, Code 660,
Greenbelt, MD  20771; turner@lucretia.gsfc.nasa.gov.}

\begin{abstract}

There are several Seyfert galaxies for which there is a discrepancy
between the small column of neutral hydrogen deduced from X-ray
observations and the much greater column derived from the reddening of the
optical/UV emission lines and continuum. The standard paradigm
has the dust within the highly ionized gas which produces
O~VII and O~VIII absorption edges (i.e., a ``dusty warm absorber''). 
We present an alternative model in which the dust exists in
a component of gas in which hydrogen has been stripped,
but which is at too low an ionization state to possess significant
columns of O~VII and O~VIII (i.e, a ``lukewarm absorber''). The lukewarm
absorber is at sufficient radial
distance to encompass much of the narrow emission-line region, and thus
accounts for the narrow-line reddening, unlike the dusty warm 
absorber. We test the model by using a combination of photoionization 
models and absorption edge fits to analyze the combined
{\it ROSAT}/{\it ASCA} dataset for the Seyfert 1.5 galaxy, NGC 3227. 
We show that the data are well fit by a combination of 
the lukewarm absorber and a more highly ionized component similar
to that suggested in earlier studies. We predict that the lukewarm
absorber will produce strong 
UV absorption lines of N~V, C~IV, Si~IV and Mg~II.
Finally, these results illustrate that singly 
ionized helium is an important, and often overlooked, source of opacity in the
soft X-ray band (100 - 500 eV).

\end{abstract}


\section{Introduction}

The presence of absorption edges of O~VII and O~VIII in
the X-ray (Reynolds 1997; George et al. 1998a) indicates that there
is a significant amount of intrinsic ionized material along our line-of-sight 
to 
the nucleus in
a large fraction ($\sim$ 0.5) of Seyfert 1 galaxies.
In addition to highly ionized gas (referred to as an X-ray or ``warm'' absorber), 
X-ray spectra often show evidence for a less-ionized
absorber. This component has been modeled using neutral gas 
(cf. George et al. 1998b), and its relationship to the 
``warm'' absorber is unclear.
Interestingly, there are several instances in which this additional neutral 
column is too small by as much as an order of magnitude to explain
the reddening of the continuum and emission lines, assuming 
typical Galactic dust/gas ratios (cf. Shull \& van Steenburg 1985). 
This inconsistency was first noted
in regard to the absence of
high ionization emission lines in the {\it IUE} spectra of MCG -6-30-15 (Reynolds
\& Fabian 1995). 
The first quantitative comparison of the neutral
columns inferred from the X-ray data to that derived from the reddening was 
for the QSO IRAS 13349+2438 by Brandt, Fabian, \& Pounds (1996), who
suggested that the dust exists within the highly ionized
X-ray absorber (ergo, a dusty warm absorber). 
It has been suggested
 that dusty warm absorbers are present in several other Seyferts 
(NGC 3227: Komossa \& Fink 1997a;
NGC 3786: Komossa \& Fink 1997b; IRAS 17020+4544: Leighly et al. 1997,
Komossa \& Bade 1998; MCG -6-30-15: Reynolds et al. 1997). 
Since it is unlikely that dust could form
within the highly ionized gas responsible for the O~VII and O~VIII
absorption, it has been suggested that
the dust is evaporated off the putative molecular torus (at $\sim$ 1 pc) and, 
subsequently, swept up in an radially outflowing wind (cf. Reynolds 1997).

In this paper, we present an alternative explanation.
It is possible that there is a component 
of dusty gas (which we will refer to as the ``lukewarm'' absorber), with an 
ionization state such that hydrogen is nearly completely ionized but
the O~VII and O~VIII columns are negligible, which has
a sufficient total column to
account for the reddening. Such a possibility has been mentioned by
Brandt et al. (1996), while Reynolds et al. (1997) have suggested that
the dusty warm absorber in MCG -6-30-15 may have multiple zones. 
Here we suggest that the lukewarm absorber lies far into the 
narrow-line region (NLR).
Such a component has been detected 
in the Seyfert galaxy NGC 4151 (Kraemer et al. 1999), and it
lies at sufficient radial distance to cover much of the NLR.
We will demonstrate that the combination of a 
dusty lukewarm absorber and a more highly ionized (O~VII and O~VIII)
absorber is consistent with the observed X-ray data
and with the 
reddening of the narrow emission lines in the Seyfert 1 galaxy NGC 3227.

\section{Absorption and Reddening in NGC 3227}

 NGC 3227 ($z$ $=$ 0.003) is a well studied Sb galaxy with an active nucleus,
 usually classified as a Seyfert 1.5 (Osterbrock \& Martel 1993). 
 X-ray observations of NGC 3227 with {\it ROSAT}
 and {\it ASCA} reveal the presence of ionized
 gas along the line-of-sight to the nucleus (Ptak et al. 1994; 
 Reynolds 1997; George et al. 1998b). 
 Using the combined {\it ASCA} and {\it ROSAT} dataset
 obtained in 1993, 
 George et al. 
 characterized the absorber with an ionization
 parameter (number of photons with energies $\geq$ 1 Ryd
 per hydrogen atom at the ionized
 face of the cloud) U $\approx$ 2.4, and a column density 
 N$_{\rm H}$ $\approx$ 3 x 10$^{21}$ cm$^{-2}$.

 The UV and optical emission lines and continuum in NGC 3227 are heavily 
 reddened. 
 Cohen (1983) measured a narrow H$\alpha$/H$\beta$ ratio $\approx$ 4.68,
 and derived 
 a reddening of E$_{B-V}$ $=$ 0.51 $\pm$ 0.04, assuming the intrinsic
 decrement to be equal to the Case B value (Osterbrock 1989).
 Cohen derived a somewhat larger reddening from the [S~II] lines, E$_{B-V}$ 
 $=$ 0.94 $\pm$0.23, which may be less reliable due to the weakness of 
 [S~II] $\lambda$4072  
(see Wampler 1968).
 The ratio of 
 broad H$\alpha$/H$\beta$ $\approx$ 5.1, indicating that the broad and narrow 
 lines are similarly reddened. Winge et al. (1995) used the
 total (narrow $+$ broad) H$\alpha$/H$\beta$ ratio to derive a somewhat 
 smaller reddening, E$_{B-V}$ $\approx$ 0.28.
 {\it IUE} spectra show the UV 
 continuum of NGC 3227
 is also heavily reddened (Komossa \& Fink 1997a). Based on the Balmer lines, 
 assuming Galactic dust properties 
 and dust/gas ratio, the derived reddening requires a hydrogen 
 (H~I and H~II combined)
 column density $\geq$ 2 x 10$^{21}$ cm$^{-2}$
 (cf. Shull \& Van Steenberg 1985), which
 is much greater than the estimated neutral column,
 but similar to that of the ionized gas 
 detected in X-rays in 1993 
 ($N_{\rm H}$ $\approx$ 3 x 10$^{21}$ cm$^{-2}$; 
 George et al. 1998b).
  
Several workers have suggested that NGC~3227 contains a screen of 
neutral material (in addition to that in the Galaxy)
along the line-of-sight to the nucleus.
Besides the ionized absorber, 
both Komossa \& Fink (1997a) and George et al. (1998b) 
found a column density of $\lesssim$ 3 x 10$^{20}$ cm$^{-2}$
of neutral material, in addition to
the Galactic column ($\sim$ 2.1 x 10$^{20}$ cm$^{-2}$, cf. Murphy et al.
1996), is required to model the X-ray spectrum 
below $\sim$500~eV.
A higher column density ($\sim$ 6 x 10$^{20}$ cm$^{-2}$)
has been suggested based on 21-cm VLA observations 
(Mundell et al. 1997).
However the angular resolution of the VLA data is poor 
(12$''$, or $\sim$ 850 pc) and Mundell et al. did not make a direct detection 
of H~I absorption against the radio continuum source in the inner nucleus.
In this paper we argue that based on the current data, there is
no reason to include a significant column of completely neutral material. 
We show that a dusty lukewarm absorber lying outside the NLR
is consistent with both the reddening of the optical continuum and 
narrow lines, and with the attenuation of the X-ray spectrum 
below $\sim$500~eV.

\section{Modeling The Absorber}

\subsection{The Lukewarm Component}

 The photoionization code we use has been described in previous publications
 (cf. Kraemer et al. 1994). For the sake of simplicity, we assume that the
 lukewarm absorber can be represented as a single zone, described by
 one set of initial conditions (i.e., density, ionization parameter, 
 elemental abundances, and dust fraction). The gas is ionized by
 the continuum radiation emitted by the central source in the active
 nucleus of NGC 3227.

 In order to fit the SED, we first determined the intrinsic luminosity at
 the Lyman limit. Since NGC 3227 is heavily reddened, we fit the
 value at the Lyman limit based on the optical continuum flux.
 From the average fluxes measured by Winge et al. (1995), after correcting for 
 a reddening of E$_{B-V}$ $=$ 0.4 (the average of the reddening
 quoted by Cohen (1983) and Winge et al.), assuming
 the reddening curve of Savage \& Mathis (1979), we find that
 the intrinsic flux at 5525 \AA~ is 
 F$_{\lambda}$ $\approx$ 2.5 x 10$^{-14}$
 ergs s$^{-1}$ cm$^{-2}$ \AA$^{-1}$.
 Interestingly, the optical luminosities of NGC 3227 and NGC 4151
 are roughly equal and, therefore, we have made the assumption that the
 two galaxies have similar optical-UV SEDs. Using the same ratio of
 optical to UV flux for NGC 3227 as in NGC 4151 (Nelson et al. 1999), we 
 determine
 F$_{\nu}$ at the Lyman limit to be 
 $\sim$ 6.2 x 10$^{-26}$ ergs s$^{-1}$ cm$^{-2}$ Hz$^{-1}$. The
 X-ray continuum from 2 -- 10 keV can be fit with an index,
 $\alpha$ $\approx$ 0.6 (George et al. 1998b), from which
 we derive a flux at 2 keV of $\sim$ 2.20 x 10$^{-29}$ ergs s$^{-1}$ cm$^{-2}$ 
 Hz$^{-1}$. Since an extrapolation 
 of the X-ray continuum
 underpredicts the
 flux at the Lyman limit by more than two orders of magnitude, the continuum
 must steepen below 2 keV. 
Hence, we have modeled the 
 EUV to X-ray SED 
as a series of power-laws of the form
 F$_{\nu}$ $\propto$ $\nu^{-\alpha}$, 
with
$\alpha = 1$ below 13.6~eV, 
$\alpha = 2$ over the range 13.6~eV $\leq h\nu <$ 500~eV, 
and $\alpha = 0.6$ above 500~eV.

Given this simple parameterization, the 
steepening of the continuum cannot occur at a much lower energy, otherwise
the EUV continuum would be too soft\footnote[2]{It is possible to have a lower break energy and a 
sufficient number of He~II ionizing photons if the EUV continuum has
a significant ``Big Blue Bump'', as suggested by Mathews \& Ferland (1987).
Although assuming such a continuum does not appreciably affect our predictions,
a full exploration of parameter space is beyond the scope of this paper.}
 to produce the observed He~II $\lambda$4686/H$\beta$ ratio
($\approx$ 0.23), specifically since the strong [O~I] $\lambda$6300
line indicates that much of the NLR gas is optically thick (for the
relative narrow emission-line strengths, see Cohen 1983).
The luminosity in ionizing photons, from 13.6 -- 10,000 eV, is
$\sim$ 1.5 x 10$^{53}$ photons s$^{-1}$. 

We have assumed roughly solar element abundances (cf. Grevesse \& Anders 1989),
which are, by number relative to H, as follows: He $=$ 0.1, 
C $=$ 3.4 x 10$^{-4}$, N $=$ 1.2 x 10$^{-4}$, O $=$ 6.8 x 10$^{-4}$,
Ne $=$ 1.1 x 10$^{-4}$, Mg $=$ 3.3 x 10$^{-5}$, Si $=$ 3.1 x 10$^{-5}$,
S $=$ 1.5 x 10$^{-5}$, and Fe $=$ 4.0 x 10$^{-5}$.
 We assume that both
silicate and carbon dust grains are present in the gas, with a power-law
distribution in sizes (see Mathis, Rumpl, \& Nordsieck 1977). Thus, we
have modified the abundances listed above by depletion
of elements from gas-phase onto dust grains, as follows (cf. Snow \& Witt
1996): C, 20\%; O, 15\%; Si, Mg and Fe, 50\%. 

For our model, we require that 1) the absorber lies outside the
majority of the NLR emission, and 2) the column of gas is fixed
to obtain the observed reddening.
Based on the  WFPC2 narrow-band [O~III] $\lambda$5007 imaging
(Schmitt \& Kinney 1996), we have
placed the lukewarm absorber at least 100 pc from the central source, 
and truncated the model at a hydrogen column density N$_{H}$ 
$=$ 2 x 10$^{21}$ cm$^{-2}$. We
adjusted the ionization
parameter such that the model 
produced a reasonable match to the absorption in the 
observed soft X-ray continuum (see below).

\subsection{Comparison to the X-ray data}

To compare our model predictions to the X-ray
data,
we used the 1993 
{\it ASCA} (0.6--10~keV) and {\it ROSAT} PSPC (0.1--2.5~keV)
data described in George et al (1998b), excluding the flare 
(``t3'' in Fig 4 of  George et al). Following standard 
practice, the normalizations of each of the 4 {\it ASCA} instruments 
and of the {\it ROSAT} dataset were allowed to vary independently.
The 5-7~keV band was also excluded from the analysis due to the 
presence of the strong, broad Fe emission line.
We assumed the continuum described above, except that the 
spectral index above 500~eV was allowed to vary during the analysis.
In addition to the lukewarm absorber (and Galactic absorption),
any highly-ionized gas was modeled by a series of 
edges of fixed energy.
This is somewhat problematic since neither the {\it ASCA} nor 
{\it ROSAT} instruments have sufficient 
spectral resolution and/or sensitivity to resolve all the possible 
edges. We have therefore 
tested for the edges most likely to be visible in highly ionized gas
(e.g., O~VII and O~VIII).

An acceptable fit to the data was obtained 
($\chi^2 = 1195$ for 1176 degrees of freedom; $\chi^2_{\nu} = 1.02$) 
with the following parameters for the lukewarm absorber:
U $=$ 0.13, n$_{H}$ $=$ 20 cm$^{-3}$, and the distance of the cloud from the
ionizing source is $\approx$ 120 pc. The predicted electron temperature
at the ionized face of this component is $\approx$ 18,000K and, therefore, it is thermally stable
(cf. Krolik, McKee, \& Tarter 1981).
The best-fitting value for the spectral index above 500~eV 
was $0.58\pm0.03$, consistent with 
our initial assumptions. We find evidence of absorption by several ions,
with the following column densities: C~V, 6.5 x 10$^{17}$ cm$^{-2}$; O~VI,
2.9 x 10$^{17}$ cm$^{-2}$; O~VII, 7.8 x 10$^{17}$ cm$^{-2}$;
O~VIII, 1.0 x 10$^{18}$ cm$^{-2}$; and, Ne~IX, 4.9 x 10$^{17}$ cm$^{-2}$. The 
O~VII and O~VIII edges translate to an effective 
hydrogen column density of highly-ionized gas of
$\gtrsim$2 x 10$^{21}$ cm$^{-2}$. 
The data/model ratios from this fit are shown in Fig. 1. The slight
underprediction of the absorption below 300 eV in the {\it ROSAT} band
is easily rectified by a small ($\sim$ 20\%) increase in the column density of the
lukewarm absorber.

The ionic column densities 
for the lukewarm absorber are listed in Table~1
and, as expected, the column densities of O~VII and
are too small to make detectable 
contributions to the X-ray absorption edges ($\tau_{\rm OVII}$ $<$ 0.01,
as opposed to $\approx$ 0.19 for the highly-ionized gas). Therefore, this 
component does not resemble 
the X-ray absorbers most frequently discussed to date
(eg Reynolds 1997; George et al 1998a).
On the other hand, the model
predicts substantial columns for H~I, N~V, Si~IV, C~IV, and Mg~II, which
would result in strong and, for the most part, saturated, UV resonance 
absorption lines.

{\it ASCA} observed NGC~3227 again in 1995. 
George et al (1998b) have shown that 
during this epoch the observed spectrum was significantly different, 
consistent with a thick, highly-ionized cloud moving into and 
attenuating $\sim$85\% of the line-of-sight to X-ray source.
Under our hypothesis that the lukewarm absorber is located 
$\gtrsim$100~pc from the nucleus, hence, we do not expect 
it to have varied between these two epochs. 
We have therefore checked and found that indeed the 1995 {\it ASCA} 
dataset are consistent with 
the soft X-ray attenuation from our lukewarm absorber. 
(Although the lack of simultaneous {\it ROSAT} PSPC data 
during 1995 prevents a stringent test.)

\section{Discussion}

We have shown that the X-ray spectrum of NGC~3227 is 
consistent with attenuation by the sum of a highly-ionized absorber 
and a lukewarm absorber.
We suggest that these are physically different components of the 
circumnuclear material surrounding NGC~3227. 
The characteristics of the  highly-ionized absorber
are similar to those previously suggested for NGC 3227
(George et al. 1998b);  
this is in the range of 
and generally similar to those in other Seyfert 1s (Reynolds
1997; George et al. 1998a). 
Such absorbers have been observed to vary on timescales 
$\lesssim$3~yr (and much faster in some cases), and are probably 
due to gas well within the NLR.

The main result of this paper is that the second component, 
our ``lukewarm absorber'', has the appropriate physical conditions 
to simultaneously explain the absorption seen 
below 0.5 keV in the X-ray band (previously modeled as 
completely neutral gas) {\it and} the 
reddening seen in the optical/UV.
Agreement with the soft X-ray data is the result of 
the lukewarm gas containing significant opacity due to
He~II. 
Agreement with the reddening of the narrow emission lines 
places the component outside the NLR.

Although the lukewarm absorber has the appropriate physical conditions and 
radial distance to redden the NLR, it must also have a sufficiently
high covering fraction to be detected.
For example, Reynolds (1997) found that 4/20 of radio-quiet active galaxies show both
intrinsic X-ray absorption and reddening. Thus,
the global covering factor of the dusty ionized absorber must be
20\%, within the solid angle that we see these objects (cf., Antonucci
1993). Kraemer et al. (1999) have shown that the covering factor 
for optically thin gas in NGC 4151, similar to our lukewarm model, can be quite
large ($\sim$ 30\%). In addition, Crenshaw et al. (1999) find
that $\sim$ 60\% of Seyfert 1 galaxies have UV absorbers with a 
global covering factor $\geq$ 50\%, and an ionization parameter
similar to the lukewarm absorber, but with lower columns on average
(cf. Crenshaw \& Kraemer 1999). Therefore, it is entirely plausible that there would be 
optically thin NLR gas along our line-of-sight to the nucleus in a fraction
of Seyfert 1s. 

The lukewarm model predicts a column of Mg~II of 3.3 x 10$^{14}$ cm$^{-2}$,
which would produce strong Mg~II $\lambda$2800 absorption.
It is interesting that NGC 3227 is one of the few Seyfert 1s to show
evidence of Mg~II $\lambda$2800 in absorption (Ulrich 1988). While
Kriss (1998) has shown that Mg~II absorption can arise in clouds 
characterized by small
column density and low ionization parameter (N$_{H}$ $\sim$ 10$^{19.5}$ 
cm$^{-2}$, U $\sim$ 10$^{-2.5}$), our results predict that it may also
arise in a large column of highly ionized NLR gas, even if a substantial
fraction of Mg is depleted onto dust grains. 

The lukewarm model predicts average grain temperatures of 30K -- 60K,
for grains with radii from 0.25 $\mu$m -- 0.005 $\mu$m, respectively. The
reradiated IR continuum, which is produced primarily by the silicate
grains (cf. Mezger, Mathis, \& Panagia 1982), peaks near 60 $\mu$m. 
Assuming a covering factor of unity, this component only accounts for
$\sim$ 1\% of the observed IR flux from NGC 3227, (which is $\approx$ 7.98 Jy
at 60 $\mu$m; {\it The IRAS Point Source Catalog} [1985]). It is likely that
most of the thermal IR emission in NGC 3227 arises in the dense
(n$_{H}$ $\geq$ 10$^{3}$ cm$^{-3}$) NLR gas in which the narrow emission
lines are formed, as is the case for the Seyfert 2 galaxy, Mrk 3
(Kraemer \& Harrington 1986).

\section{Summary}

 Using the combined 1993 {\it ROSAT}/{\it ASCA} dataset for NGC 3227, 
 and photoionization model predictions, we have demonstrated that the observed
 reddening may occur in dusty, photoionized gas which is in a much lower
 ionization state than X-ray absorbers detected
 (by their O~VII and O~VIII edges)
 in Seyfert 1 galaxies (cf. Reynolds 1997; George et al. 1998a) and the
 dusty warm absorbers that have been proposed (cf. Komossa \& Fink 1997a). This 
 component (the lukewarm absorber) is $\sim$ 120 pc from the
 central ionizing source and its physical conditions are
 similar to those in optically thin gas present in the NLR of the Seyfert 1 
 galaxy NGC 4151. If this model is correct, we predict that strong UV 
resonance absorption lines with high column densities
 from the lukewarm absorber will be observed in NGC 3227.

 We have confirmed earlier results regarding the presence of an X-ray absorber
 within NGC 3227 with O~VII and O~VIII optical depths similar to those
 determined by Reynolds (1997). This component lies closer to the
 central source than the lukewarm absorber, but is essentially transparent
 to EUV and soft X-ray radiation and, hence, does not effectively screen
 the lukewarm gas. We find no requirement
 for neutral gas in addition
 to the Galactic column.

 These results illustrate that a moderately large ($\sim$ 10$^{21}$ cm$^{-2}$)
 column of ionized gas can produce significant soft X-ray absorption if
 much of the helium is in the singly ionized state. 
 Since
clouds with large He~II columns may be a common feature of 
the NLR of Seyfert galaxies, such a component should be 
included in modeling the X-ray absorption.
If such a component
 is present along our line-of-sight in an active galaxy, it is
 likely that the intrinsic neutral column has been overestimated.  

\acknowledgments

 S.B.K. and D.M.C. acknowledge support from NASA grant NAG5-4103. T.J.T.
 acknowledges support from UMBC and NASA/LTSA grant NAG5-7835.

\clearpage

\figcaption[fig1a.ps]{
{\it Upper Panel}: The spectral components described in the text. 
The SED in the EUV (dotted line) is respresented by the 3 power laws
described in \S3.1. This is attenuated by a highly-ionized absorber
(giving rise to the O VII \& O VIII edges), and a dusty 
lukewarm absorber (giving rise to the H I \& He II edges, as well as
additional opacity throughout the spectrum below $\sim$1~keV). 
Finally the spectrum is attenuated by Galactic absorption leading to the 
observed spectrum (bold line).
{\it Lower panel}: The data/model ratios for the {\it ROSAT}\ PSPC data
(open circles) and {\it ASCA}\ SIS data (filled triangles). 
The SIS data are weighted means of the 2 instruments.
Both the PSPC and SIS datasets have been rebinned in energy-space 
for clarity.
The 5--7~keV band was excluded from the analysis due to the 
intense Fe K-shell emission (open triangles).}
\label{fig1} 

%\figcaption[bbb_low.eps]{
%}\label{fig2} 

%\figcaption[sed1.eps]{
%}\label{fig3} 

%\figcaption[sed2.eps]{
%}\label{fig4} 

%\figcaption[fig5.eps]{
%}\label{fig5} 

\clearpage
\plotfiddle{fig1a.ps}{11cm}{270}{70}{70}{-280}{380}

%\clearpage
%\plotone{fig2.eps}

%\clearpage
%\plotone{fig3.eps}

%\clearpage
%\plotone{fig4.eps}

%\clearpage
%\plotone{fig5.eps}

%\clearpage
%\plotone{fig4.eps}

%\clearpage
%\plotone{fig5.eps}
\end{document}